\documentclass[pra,a4paper,nofootinbib,twocolumn,showpacs,preprintnumbers,amsmath,amssymb,floatfix,amstex,superscriptaddress]{revtex4}
\newcommand{\ket}[1]{|#1\rangle}                        %
\usepackage{graphicx,graphics,wrapfig,rotating}         
\usepackage{dcolumn}                                    
\usepackage{bm,fancybox}                                
\usepackage{times,euscript,eufrak,oldgerm}              %
\usepackage[german,english]{babel}                      %
\usepackage{psfrag}
\usepackage{ulem}
\usepackage{color}
\begin{document}
\title{Universal quantum computation in decoherence-free subspaces with hot trapped-ions} %
\author{Leandro Aolita}
\affiliation{%
Instituto de F\'\i sica, Universidade Federal do Rio de Janeiro. Caixa Postal
68528, 21941-972 Rio de Janeiro, RJ, Brazil.\\
}
\affiliation{%
Max-Planck-Institute f\"ur Physic Komplexer Systeme, N\"othnitzerstrasse 38, D-01187, Dresden, Germany\\
}
\author{Luiz Davidovich}
\affiliation{%
Instituto de F\'\i sica, Universidade Federal do Rio de Janeiro. Caixa Postal
68528, 21941-972 Rio de Janeiro, RJ, Brazil.\\
}%
\author{Kihwan Kim}
\affiliation{%
Institut f\"ur
Experimentalphysik, Universit\"{a}t Innsbruck, Technikerstra{\ss}e 25, A-6020
Innsbruck, Austria\\
}%
\author{Hartmut H\"{a}ffner}
\affiliation{%
Institut f\"ur
Quantenoptik und Quanteninformation der \"Osterreichischen Akademie der
Wissenschaften, Technikerstra{\ss}e 21a, A-6020 Innsbruck, Austria\\
}%
\date{\today}%
\newcommand{\newnew}[1]{\textcolor{green}{#1}}
\newcommand{\Leandro}[1]{\textcolor{green}{\it{#1}}}
\newcommand{\new}[1]{\textcolor{red}{#1}}
\newcommand{\old}[1]{\textcolor{blue}{\sout{#1}}}
\begin{abstract}
We consider interactions that
generate a universal set of quantum gates on logical qubits encoded
in a collective-dephasing-free subspace, and discuss their
implementations with trapped ions. This allows for the removal of the by-far largest source of decoherence in current trapped-ion experiments, collective dephasing. In addition, an explicit parametrization of all two-body
Hamiltonians able to generate such gates without the system's
state ever exiting the protected subspace is provided.
\end{abstract}
\maketitle
\section{Introduction}
In quantum information processing tasks decoherence can be overcome
either by an active approach or by a passive one. The former
consists, in analogy with classical computation, of encoding
information in a redundant fashion by means of the so-called
error-correcting codes. In this approach information is encoded in
subspaces of the total Hilbert space of the system in such a way that ``errors''
induced by the interaction with the environment can be detected and
corrected without gaining information about the actual state of the
system prior to corruption \cite{Shor-Machiavello-Steane-Gottesman-Calderbank-Miquel}.
\par The passive approach, on the other hand, is an error preventing
scheme, in which logical qubits are encoded within decoherence-free subspaces (DFS), which do not decohere because  of
symmetry \cite{Palma-Barenco-Zanardi}. A simple example is provided
by a system of $N$ spins collectively interacting with the same
reservoir, for which the interaction is mediated by the collective
angular momentum raising and lowering operators
$\hat{S}^{+}\equiv\sum_{i=1}^{N}\hat{S}^{+}_{i}$ and
$\hat{S}^{-}\equiv\sum_{i=1}^{N}\hat{S}^{-}_{i}$, where
$\hat{S}^{+}_{i}$ and $\hat{S}^{-}_{i}$ are the corresponding
raising and lowering operators, respectively, of the $i$-th
particle. The collective operators have no support on the
eigenstates of the total squared angular momentum $\hat{S}^2$
corresponding to zero eigenvalue. The evolution of these eigenstates
is therefore unitary because they simply do not couple to the
reservoir; and they can be used as a logical-qubit basis for
decoherence-free quantum computation \cite{Lidar-Bacon, Bacon}.
\par When the coupling to the environment is mediated by the collective $z$-angular-momentum operator $\hat{S}^{z}\equiv\sum_{i=1}^{N}\hat{S}^{z}_{i}$, the type of noise is called {\it collective dephasing}. The interaction Hamiltonian between the system and the bath is then proportional to $\hat{S}^{z}\otimes\hat{B}$, where
$\hat{B}$ is an arbitrary operator acting on the Hilbert space
associated to the bath. The action of this type of bath  is
equivalent to that of randomly-fluctuating fields: a general
qubit-state $\ket{\Psi}\equiv a\ket{0}+b\ket{1}$ transforms as
$\ket{\Psi}\rightarrow a\ket{0}+be^{i\zeta}\ket{1}$, which leads to
the loss of coherence of the state for $\zeta$ is a random
fluctuating phase. By using one pair of physical qubits, whose
members are labeled by the subindexes $i_{1}$ and $i_{2}$, to encode
logical qubit $i$, one can protect information from the detrimental
action of decoherence. In fact, the well-known
\cite{Duan-Zanardi-Lidar1,Kielpinsky2} logical basis
$B_{L_{i}}\equiv\{\ket{0_{L_{i}}}\equiv\ket{0_{i_{1}}1_{i_{2}}};\ket{1_{L_{i}}}\equiv
\ket{1_{i_{1}}0_{i_{2}}}\}$ spans a DFS protected against
collective dephasing, which we call $\mathbb{V}_{DFS_{2_{i}}}$. That
is, the logical state $\ket{\Psi_{L_{i}}} \equiv
a\ket{0_{L_{i}}}+b\ket{1_{L_{i}}}=a\ket{0_{i_{1}}}\ket{1_{i_{2}}}+b\ket{1_{i_{1}}}\ket{0_{i_{2}}}$
evolves as $\ket{\Psi_{L_{i}}} \rightarrow
a\ket{0_{i_{1}}}e^{i\zeta}\ket{1_{i_{2}}}+be^{i\zeta}\ket{1_{i_{1}}}\ket{0_{i_{2}}}
= e^{i\zeta}(a\ket{0_{L_{i}}}+b\ket{1_{L_{i}}})$ and is thus
invariant  up to an irrelevant global phase factor.
\par Two pairs of physical qubits, whose members are labeled by
the subindexes $i_{1}$ and $i_{2}$, and $j_{1}$ and $j_{2}$,
respectively, are in turn needed to encode two logical qubits $i$
and $j$ . The direct product subspace
$\mathbb{V}_{DFS_{2_{i}}}\otimes\mathbb{V}_{DFS_{2_{j}}}$, spanned by
the basis $B_{L_{i}}\otimes B_{L_{j}}$, yields a DFS. However, one
should note that this is not the total protected subspace supported
by all four qubits if all four physical qubits experience the
same phase fluctuations.  In this case the states
$|0_{i_{1}}0_{i_{2}}1_{j_{1}}1_{j_{2}}\rangle$ and
$|1_{i_{1}}1_{i_{2}}0_{j_{1}}0_{j_{2}}\rangle$, which are outside
$\mathbb{V}_{DFS_{2_{i}}}\otimes\mathbb{V}_{DFS_{2_{j}}}$, are also
protected against collective dephasing for they have the same amount
of excitations as the states in
$\mathbb{V}_{DFS_{2_{i}}}\otimes\mathbb{V}_{DFS_{2_{j}}}$. In
general, any coherent superposition of states with the same amount
of excitations is immune against collective dephasing. Thus, the
total protected subspace, which we call $\mathbb{V}_{DFS_{4_{ij}}}$,
is that spanned by $B_{L_{i}}\otimes B_{L_{j}}$ together with the
states $|0_{i_{1}}0_{i_{2}}1_{j_{1}}1_{j_{2}}\rangle$ and
$|1_{i_{1}}1_{i_{2}}0_{j_{1}}0_{j_{2}}\rangle$. If pairs $i$ and $j$ are further apart than the typical noise correlation length ---but with both
qubits from each pair still subject to to the same fluctuations--- $\mathbb{V}_{DFS_{2_{i}}}\otimes\mathbb{V}_{DFS_{2_{j}}}$ is the only protected
subspace.
\par On the experimental side, demonstration of immunity of a DFS of two
photons to collective noise was accomplished in \cite{Kwiat} and
realizations of DFS's for nuclear magnetic resonance (NMR) systems
were carried out in \cite{Fortunato-Viola}. Demonstration of a
collective-dephasing-free quantum memory of one logical qubit
composed of a pair of trapped $^{9}Be^{+}$ ions was first achieved
in \cite{Kielpinsky} and coherent oscillations between two logical
states, encoded into the two Bell states
$\ket{\Psi_{\pm}}=\frac{1}{\sqrt{2}}(\ket{01}\pm\ket{10})$, by
inducing a gradient of the magnetic field applied to both ions, were
reported in \cite{Roos, Langer}. Finally, entanglement lifetimes of
more than 7 seconds \cite{Langer} and robust entanglement lasting
for more than 20 seconds \cite{Haffner2} were attained using ground
state hyperfine levels of $^{9}Be^{+}$ ions and ground state Zeeman
sublevels of $^{40}Ca^{+}$ ions, respectively. These experiments
demonstrated that for trapped-ions collective-dephasing is the
major source of qubit decoherence. We therefore focus on this type
of noise throughout the rest of the paper. Nevertheless, apart from
the proof-of-principle experiments mentioned above, demonstrating
the robustness of these subspaces, experimentally accessible
implementations of DFS-encoded gates are still sparse; and in spite
of the rich (but abstract) body of work  on DFS's, a universal set
of gates between two encoded logical qubits is yet to be
demonstrated.
\par Proposals for ion trap quantum computing with DFS's exist, and
they are essentially divided into two families that complement each
other. In the first paradigm \cite{Kielpinsky2} gates between two
logical qubits are implemented \cite{Kielpinsky2, Wu-Zanardi-Lidar}
by bringing together two pairs of ions (each pair encoding a logical
qubit), initially stored in memory regions, to an interaction region
where a simultaneous interaction among all four ions takes place
according to the S\o rensen-M\o lmer (SM) gate described in
\cite{Sorensen-Molmer, Sackett}. Individual laser addressing is not
necessary for this scheme, but a reliable ion-shuttling technique is
an essential requirement. In addition, even though this scheme maps
$\mathbb{V}_{DFS_{4_{ij}}}$ into itself, it does not preserve the
state inside the DFS throughout the gate evolution \cite{Bacon}. The
second paradigm \cite{Duan, Cen-Wang-Wang} works in the individual
laser addressing regime and relaxes the need of ion-shuttling. In
this approach, ions are trapped in a crystal-like effective
potential created by arrays of multi-connected linear Paul traps.
Each ion is associated to a neighbor to form a pair that encodes one
logical qubit \cite{Cen-Wang-Wang}. By inducing a
$\hat{\sigma}^{z}$-dependent force (see \cite{Leibfried, Lee,
Blinov} and references therein) on two ions, each from different
pairs, it is in principle possible to implement a (geometric phase)
$\hat{\sigma}^{z}$-gate between the logical qubits encoded into both
pairs. Particular advantages of these $\hat{\sigma}^{z}$-gates are
that they can be considerably fast and robust. It has been
conjectured \cite{Lee,Blinov,Haljan} though, that these
$\hat{\sigma}^{z}$-gates are  very ineffecient with
magnetic-field-insensitive (or ``clock'') states, which possess such
remarkable coherence properties \cite{Langer, Clock-states}.
However, it would be very advantageous to combine clock states with  DFSs as this would lead to very long coherence times and minimize the overhead due to quantum error correction.
\par In our present paper we assess different possible interactions involving only two physical qubits at a time that generate
universal quantum gates on DFS-encoded qubits, and describe feasible
experimental demonstrations of each of them with trapped-ions. The
work is conceptually divided into two parts. The first one (Sec.
\ref{SecII}) is devoted to the general formal classification of all
two-body dynamics able to generate universal quantum gates inside
the DFS without the system's state ever leaving it. The aim here is
not to establish the set of formal conditions for a given
Hamiltonian to generate universal DFS quantum computation, as in
\cite{Lidar-Bacon, Bacon}; but rather to explicitly construct the
allowed Hamiltonians in a simple way in terms of the Pauli operators
associated to each physical qubit. This is to serve as a simple
``classification table" for experimentalists to rapidly check
whether the type of interactions present in their given system
qualifies as a candidate for generating universal DFS quantum
computation or not. In particular, we introduce  the most general two body
Hamiltonian that generates universal quantum computation while guaranteeing 
the evolution to take place entirely 
inside $\mathbb{V}_{DFS_{2_{i}}}\otimes\mathbb{V}_{DFS_{2_{j}}}$. Furthermore, we show
that the only possible interaction between two logcial qubits which obeys the previous assumptions  is of the
type $\hat{\sigma}^{z}\otimes\hat{\sigma}^{z}$.  For the cases where
leakage out of
$\mathbb{V}_{DFS_{2_{i}}}\otimes\mathbb{V}_{DFS_{2_{j}}}$  into
$\mathbb{V}_{DFS_{4_{ij}}}$ is allowed, we consider the encoding
re-coupling scheme originally introduced in \cite{Lidar-Wu3} for NMR
systems. There, a maximally entangling gate is implemented on the
DFS through a sequence of transformations that momentarily takes the
composite state out of
$\mathbb{V}_{DFS_{2_{i}}}\otimes\mathbb{V}_{DFS_{2_{j}}}$ but never
out of $\mathbb{V}_{DFS_{4_{ij}}}$.
\par The second part (Sec. \ref{SecIII}) describes the technical details of the implementation on trapped-ions of the ideas
presented in Sec.~\ref{SecII}. Our implementations work in the
individual laser addressing regime and require no ion-shuttling. We
show that for the realization of local and conditional gates
inside $\mathbb{V}_{DFS_{2_{i}}}\otimes\mathbb{V}_{DFS_{2_{j}}}$,
the SM-gate and the $\hat{\sigma}^{z}$-gate, respectively, can be
used. For the realization of the encoded re-coupling scheme in turn,
an alternative two-physical-qubit gate is required.  The latter is
based on bichromatic Raman fields and applies to all states in
general regardless of their magnetic properties, including clock
states connected via dipole Raman transitions. Furthermore, this
gate does not require the ions to be in their motional ground state,
provided that they always remain in the Lamb-Dicke regime.
Therefore, it is a potentially useful alternative to the SM-gate and
the $\hat{\sigma}^{z}$-gate also
outside the context of DFS's. Our conclusions are finally summarized
in section \ref{conclusion}.
\section{General Hamiltonians for universal quantum computation in the DFS}
\label{SecII}
\subsection{Local operations: the logical SU(2) Lie Algebra}
\label{logical}
We want to find a complete set of orthogonal operators mapping $\mathbb{V}_{DFS_{2_{i}}}$  (for any $i$) onto itself. We define then logical identity and Pauli operators, $\hat{\sigma}^{0}_{L_{i}}\equiv\hat{I}_{L_{i}}$, $\hat{\sigma}^{1}_{L_{i}}\equiv\hat{\sigma}^{x}_{L_{i}}$,
$\hat{\sigma}^{2}_{L_{i}}\equiv\hat{\sigma}^{y}_{L_{i}}$
and $\hat{\sigma}^{3}_{L_{i}}\equiv\hat{\sigma}^{z}_{L_{i}}$ of the $i$-th logical qubit, as:
\begin{eqnarray}
\hat{\sigma}^{0}_{L_{i}}&\equiv&\alpha_{i}\hat{\sigma}^{0}_{i_{1}}\otimes\hat{\sigma}^{0}_{i_{2}}-(1-\alpha_{i})\hat{\sigma}^{3}_{i_{1}}\otimes\hat{\sigma}^{3}_{i_{2}}+\hat{0}_{L_{i}},\nonumber\\
\hat{\sigma}^{1}_{L_{i}}&\equiv&\beta_{i}\hat{\sigma}^{1}_{i_{1}}\otimes\hat{\sigma}^{1}_{i_{2}}+(1-\beta_{i})\hat{\sigma}^{2}_{i_{1}}\otimes\hat{\sigma}^{2}_{i_{2}}+\hat{0}_{L_{i}},\nonumber\\
\hat{\sigma}^{2}_{L_{i}}&\equiv&\gamma_{i}\hat{\sigma}^{2}_{i_{1}}\otimes\hat{\sigma}^{1}_{i_{2}}-(1-\gamma_{i})\hat{\sigma}^{1}_{i}\otimes\hat{\sigma}^{2}_{i_{2}}+\hat{0}_{L_{i}},\nonumber\\
\hat{\sigma}^{3}_{L_{i}}&\equiv&\varepsilon_{i}\hat{\sigma}^{3}_{i_{1}}\otimes\hat{\sigma}^{0}_{i_{2}}-(1-\varepsilon_{i})\hat{\sigma}^{0}_{i_{1}}\otimes\hat{\sigma}^{3}_{i_{2}}+\hat{0}_{L_{i}};
\label{logPaulimat}
\end{eqnarray}
where $\hat{\sigma}^{p}_{i_{n}}$ is the identity ($p=0$) or Pauli ($1\le p\le 3$) operator associated to the $n$-th ($n=1$ or 2) physical qubit of the $i$-th pair, and with $\alpha_{i}, \beta_{i}, \gamma_{i}$, and $\varepsilon_{i}$ any real
numbers such that $0\le\alpha_{i}, \beta_{i}, \gamma_{i}$ and $\varepsilon_{i}\le 1$. The operator $\hat{0}_{L_{i}}$ represents the logical null operator, which is defined as any operator without support on $\mathbb{V}_{DFS_{2_{i}}}$. The operators in Eq. (\ref{logPaulimat}) map $\mathbb{V}_{DFS_{2_{i}}}$ onto itself and their  action on $B_{L_{i}}$ is exactly equivalent to that of the usual identity and Pauli physical operators on the computational basis.
It can be seen that Eq. (\ref{logPaulimat}) is the most general way to construct them from the operators that act on the physical qubits. For example, if we added the term $\hat{\sigma}^{1}_{i_{1}}\otimes\hat{\sigma}^{3}_{i_{2}}$ to the definition of $\hat{\sigma}^{3}_{L_{i}}$ in (\ref{logPaulimat}) we would exit $\mathbb{V}_{DFS_{2_{i}}}$; terms as $\hat{\sigma}^{1}_{i_{1}}\otimes\hat{\sigma}^{1}_{i_{2}}$ would not take us out of the DFS but act like $\hat{\sigma}^{1}_{L_{i}}$ instead; and so on. Combinations as $\hat{\sigma}^{1}_{i_{1}}\otimes\hat{\sigma}^{1}_{i_{2}}-\hat{\sigma}^{2}_{i_{1}}\otimes\hat{\sigma}^{2}_{i_{2}}$ or $\hat{\sigma}^{1}_{i_{1}}\otimes\hat{\sigma}^{3}_{i_{2}}-i\hat{\sigma}^{2}_{i_{1}}\otimes\hat{\sigma}^{0}_{i_{2}}$ are allowed though, since they have no support on $\mathbb{V}_{DFS_{2_{i}}}$ and can therefore be grouped inside  $\hat{0}_{L_{i}}$. In general there are sixteen possible products between $\hat{\sigma}^{0}_{i_{1}}$, $\hat{\sigma}^{1}_{i_{1}}$,
$\hat{\sigma}^{2}_{i_{1}}$ and $\hat{\sigma}^{3}_{i_{1}}$, and  $\hat{\sigma}^{0}_{i_{2}}$, $\hat{\sigma}^{1}_{i_{2}}$,
$\hat{\sigma}^{2}_{i_{2}}$ and $\hat{\sigma}^{3}_{i_{2}}$. Each of these products, or combinations of them, apart from those already considered in Eq. (\ref{logPaulimat}), either takes the state out of $\mathbb{V}_{DFS_{2_{i}}}$, does not have the desired action, or has no support on $\mathbb{V}_{DFS_{2_{i}}}$ and is therefore absorbed inside the definition of $\hat{0}_{L_{i}}$. The most general expression for the logical null operator is given by
\begin{widetext}
\begin{eqnarray}
\hat{0}_{L_{i}}&\equiv&
\rho_{i}(\hat{\sigma}^{0}_{i_{1}}\otimes\hat{\sigma}^{1}_{i_{2}}-i\hat{\sigma}^{3}_{i_{1}}\otimes\hat{\sigma}^{2}_{i_{2}})
+\theta_{i}(\hat{\sigma}^{1}_{i_{1}}\otimes\hat{\sigma}^{0}_{i_{2}}-i\hat{\sigma}^{2}_{i_{1}}\otimes\hat{\sigma}^{3}_{i_{2}})+
\vartheta_{i}(\hat{\sigma}^{1}_{i_{1}}\otimes\hat{\sigma}^{3}_{i_{2}}-i\hat{\sigma}^{2}_{i_{1}}\otimes\hat{\sigma}^{0}_{i_{2}})
+\zeta_{i}(\hat{\sigma}^{3}_{i_{1}}\otimes\hat{\sigma}^{1}_{i_{2}}-i\hat{\sigma}^{0}_{i_{1}}\otimes\hat{\sigma}^{2}_{i_{2}})+ \nonumber\\
&&\kappa_{i}(\hat{\sigma}^{0}_{i_{1}}\otimes\hat{\sigma}^{0}_{i_{2}}+\hat{\sigma}^{3}_{i_{1}}\otimes\hat{\sigma}^{3}_{i_{2}})
+\lambda_{i}(\hat{\sigma}^{1}_{i_{1}}\otimes\hat{\sigma}^{1}_{i_{2}}-\hat{\sigma}^{2}_{i_{1}}\otimes\hat{\sigma}^{2}_{i_{2}})+
\varsigma_{i}(\hat{\sigma}^{1}_{i_{1}}\otimes\hat{\sigma}^{2}_{i_{2}}+\hat{\sigma}^{2}_{i_{1}}\otimes\hat{\sigma}^{1}_{i_{2}})+
\xi_{i}(\hat{\sigma}^{3}_{i_{1}}\otimes\hat{\sigma}^{0}_{i_{2}}+\hat{\sigma}^{0}_{i_{1}}\otimes\hat{\sigma}^{3}_{i_{2}})\ ,
\label{lognull}
\end{eqnarray}
\end{widetext}
with $\rho_{i}, \theta_{i}, \vartheta_{i}, \zeta_{i}, \kappa_{i}, \lambda_{i}, \varsigma_{i}$, and $\xi_{i}$ any complex
numbers.
\par  The operators in Eq. (\ref{logPaulimat}) are orthonormal:
$Tr[\hat{\sigma}^{p}_{L_{i}}\hat{\sigma}^{q}_{L_{i}}]=\delta_{pq}$, with $p$ and
$q=0,1,2$ or $3$, and form therefore a complete orthonormal basis of the space of the complex operators acting on
the two-dimensional subspace $\mathbb{V}_{DFS_{2_{i}}}$. They also satisfy, inside of $\mathbb{V}_{DFS_{2_{i}}}$, the desired SU(2) usual commutation relations: $[\hat{\sigma}^{p}_{L_{i}},
\hat{\sigma}^{q}_{L_{i}}]=2i\epsilon_{pqr}\hat{\sigma}^{r}_{L_{i}}$, for $p,q$
and $r=1,2$ or $3$; and  $[\hat{\sigma}^{0}_{L_{i}}, \hat{\sigma}^{p}_{L_{i}}]=0$, for $p=0,1,2$ or $3$. As an example to
show this, we calculate explicitly the commutator $[\hat{\sigma}^{1}_{L_{i}}, \hat{\sigma}^{2}_{L_{i}}]$ and obtain
\begin{eqnarray}
[\hat{\sigma}^{1}_{L_{i}},\hat{\sigma}^{2}_{L_{i}} ] & = &
2i(1-\beta_{i}-\gamma_{i}+2\beta_{k}\gamma_{k})\hat{\sigma}
^{3}_{i_{1}}\otimes\hat{\sigma}^{0}_{i_{2}}\nonumber\\
&-&2i(\beta_{k}
+\gamma_{i}-2\beta_{i}\gamma_{i})\hat{\sigma}^{0}_{i_{1}}\otimes\hat{\sigma}^{3}_{i_{2}}\,.
\end{eqnarray}
Doing the identification
$\varepsilon'_{i}\equiv1-\beta_{i}-\gamma_{i}+2\beta_{i}\gamma_{i}$ and
since $0\le\beta_{i}\le 1$ and $0\le\gamma_{i}\le 1$ we see that
$0\le\varepsilon'_{i}\le 1$, which leads us to
\begin{equation}
[\hat{\sigma}^{1}_{L_{i}},
\hat{\sigma}^{2}_{L_{i}}]=2i\big[\varepsilon'_{k}\hat{\sigma}^{3}_{i_{1}}\otimes\hat{\sigma}^{0}_{i_{2}}-(1-\varepsilon'_{i})\hat{\sigma}^{0}_{i_{1}}\otimes\hat{\sigma}^{3}_{i_{2}}\big]\,.
\end{equation}
This is, inside of $\mathbb{V}_{DFS_{2_{i}}}$, exactly equivalent to $2i\hat{\sigma}^{3}_{L_{i}}$. Note that the logical operator
obtained here and the fourth operator in Eq.~(\ref{logPaulimat}) are actually not strictly equal, since $\varepsilon_{i}$ and $\varepsilon'_{i}$ are not
necessarily the same number. Their difference however only shows when applied to states outside $\mathbb{V}_{DFS_{2_{i}}}$, their action
on this subspace
 is exactly the same. All the other SU(2) fundamental  commutation relations are straightforwardly obtained in the same way. We see thus that
the logical Pauli operators defined in (\ref{logPaulimat}) are the most general  representation of the SU(2) Lie algebra on $\mathbb{V}_{DFS_{2_{i}}}$ constructed from the physical-qubit operators.
\par We also notice that the logical operators $\overline{X}_{i}\equiv\frac{1}{2}(\hat{\sigma}^{1}_{i_{1}}\otimes\hat{\sigma}^{1}_{i_{2}}+
\hat{\sigma}^{2}_{i_{1}}\otimes\hat{\sigma}^{2}_{i_{2}})$, $\overline{Y}_{i}\equiv\frac{1}{2}(\hat{\sigma}^{2}_{i_{1}}\otimes\hat{\sigma}^{1}_{i_{2}}-
\hat{\sigma}^{1}_{i_{1}}\otimes\hat{\sigma}^{2}_{i_{2}})$ and $\overline{Z}_{i}\equiv\frac{1}{2}(\hat{\sigma}^{3}_{i_{1}}-\hat{\sigma}^{3}_{i_{2}})$
used in \cite{Wu-Zanardi-Lidar} are a particular case of (\ref{logPaulimat}), corresponding to
$\beta_i=\gamma_i=\varepsilon_i=1/2$. These operators generate the SU(2) group on the whole Hilbert space, but they have the same
action as those defined in (\ref{logPaulimat}) on $\mathbb{V}_{DFS_{2_{i}}}$. The advantage of the logical operators in
(\ref{logPaulimat}) is that they give the experimentalist more freedom of choice, as any choice of $\alpha_{i}$, $\beta_{i}$,
$\gamma_{i}$ and $\varepsilon_{i}$ works just as well in $\mathbb{V}_{DFS_{2_{i}}}$. As a matter of fact, we exploit this freedom  below  to simplify the procedure for obtaining DFS-encoded gates for trapped-ions.
\par The situation is now completely equivalent to that of a physical qubit, with the logical states in $B_{L_{i}}$ and logical operators in
(\ref{logPaulimat}) playing the role of the physical ones. The important thing to keep in mind though is that these logical
operators allow us to operate on the logical states in the same way as their physical counterparts without ever exiting
$\mathbb{V}_{DFS_{2_{i}}}$. With this at hand we can now write down the Hamiltonian that
generates the most general unitary operation on the $i$-th logical qubit, it reads:
\begin{eqnarray}
\hat{H}_{L_{i}}\equiv
B^{0}_{i}\hat{\sigma}^{0}_{L_{i}}+B^{1}_{i}\hat{\sigma}^{1}_{L_{i}}+B^{2}_{i}
\hat{\sigma}^{2}_{L_{i}}+B^{3}_{i}\hat{\sigma}^{3}_{L_{i}},
\label{Hamlogqubit}
\end{eqnarray}
with $B^{0}_{i}$, $B^{1}_{i}$, $B^{2}_{i}$ and $B^{3}_{i}$ any real
numbers  (times arbitrary units of energy) that play the role of a
``logical magnetic field". Notice that we are explicitly including
the logical identity in Hamiltonian (\ref{Hamlogqubit}), even though
it only  introduces an irrelevant global phase factor. This is
because we want to account, in the most general way, for the
possibility of appearance of terms proportional to
$\hat{\sigma}^{3}_{i_{1}}\otimes\hat{\sigma}^{3}_{i_{2}}$, which are not irrelevant for an implementation on physical
qubits.
\subsection{Computation in $\mathbb{V}_{DFS_{2_{i}}}\otimes\mathbb{V}_{DFS_{2_{j}}}$: the two-physical-qubit interaction Hamiltonian}
\label{uniqueHam}
\par We proceed now with the interaction Hamiltonian between logical qubits $i$ and $j$, $\hat{H}_{L_{i}L_{j}}$. Under the action of
this Hamiltonian there can be no transfer of excitations between
both qubit pairs, so that each logical qubit evolves inside its own
encoded subspace. The only allowed interactions are then those ones
composed of combinations of products of logical Pauli operators of
both logical qubits. Nevertheless, the remarkable observation is
that $\hat{\sigma}^{3}_{L_{i}}$ and $\hat{\sigma}^{3}_{L_{j}}$ are
the only logical operators that do not involve interactions between
the physical qubits from the same pair. Any product of two logical
Pauli operators from both logical qubits other than
$\hat{\sigma}^{3}_{L_{i}}\otimes\hat{\sigma}^{3}_{L_{j}}$ will
necessarily contain products of more than two physical-qubit
(non-identity) operators. We see, therefore, that there exists only
one type of  two-body interaction able to generate non-trivial two
logical qubit operations  on the DFS and at the same time preserving
the composite state always inside
$\mathbb{V}_{DFS_{2_{i}}}\otimes\mathbb{V}_{DFS_{2_{j}}}$. It is
given by:
\begin{equation}
\hat{H}_{L_{i}L_{j}}\propto\hat{\sigma}^
{3}_{L_{i}}\otimes\hat{\sigma}^{3}_{L_{j}}\ .
\label{UniqueHamiltonian}
\end{equation}
This interaction between both logical qubits reduces to a simple
Ising interaction between one physical qubit from pair $i$ and one
from $j$ when the non-symmetric choice $\varepsilon_{i}$ and
$\varepsilon_{j}$ equal to $0$, or $1$, is taken. Also, the fact
that the operators in the $z$ direction play such a preferential
role is not surprising, since, for collective dephasing, it is the
total $z$ angular momentum that mediates the coupling of the qubits
to the environment; and our protected subspace is precisely that of
null total $z$ angular momentum.
\par The aim of Hamiltonians (\ref{Hamlogqubit})  and (\ref{UniqueHamiltonian}),
together with expressions (\ref{logPaulimat}) for the single
logical-qubit operators, is to provide a tool for the immediate
classification of the allowed two-body dynamics for the implementation of DFS
universal quantum computation.  Any system whose Hamiltonian cannot be expressed as
given by equations (\ref{Hamlogqubit})  and
(\ref{UniqueHamiltonian}), together with (\ref{logPaulimat}), is
automatically excluded as a candidate for such computation, except,
of course, for the possible appearance of any combination of
physical-qubit operators that can be expressed as in Eq.
(\ref{lognull}).
\subsection{Computation in $\mathbb{V}_{DFS_{4_{ij}}}$: the encoded re-coupling scheme}
\label{ERS}
\par An alternative technique to entangle logical qubits is the encoded re-coupling scheme, which was originally developed for NMR systems in \cite{Lidar-Wu3}. In this scheme, a $\hat{\sigma}^{3}\otimes\hat{\sigma}^{3}$ interaction is effectively simulated by a sequence of $\hat{\sigma}^{+}\otimes\hat{\sigma}^{-}$-type interactions between different physical qubits from both pairs. This provokes an actual transfer of excitations between both pairs, so that the logical qubits momentarily exit $\mathbb{V}_{DFS_{2_{i}}}\otimes\mathbb{V}_{DFS_{2_{j}}}$ and ``loose their encoded logical identity''. But the total amount of excitations remains the same, so that the whole evolution takes place inside $\mathbb{V}_{DFS_{4_{ij}}}$. The technique is based on the identity
\begin{eqnarray}
\label{re-coupling}
&&e^{-i[\hat{\sigma}^{+}_{i_{1}}\otimes\hat{\sigma}^{-}_{j_{1}}+h.c.]\pi/4}
e^{-i[\hat{\sigma}^{+}_{i_{1}}\otimes\hat{\sigma}^{-}_{i_{2}}+ h.
c.]\pi/2}\big[\hat{\sigma}^{+}_{i_{2}}\otimes\hat{\sigma}^{-}_{j_{1}}\nonumber
\\
&&+ h. c.\big]
e^{i[\hat{\sigma}^{+}_{i_{1}}\otimes\hat{\sigma}^{-}_{i_{2}}+ h. c.]\pi/2}
 e^{i[\hat{\sigma}^{+}_{i_{1}}\otimes\hat{\sigma}^{-}_{j_{1}}+ h. c.]\pi/4}\nonumber\\
&&=\frac{1}{2}\hat{\sigma}^{3}_{i_{2}}\otimes\big(\hat{\sigma}^{3}_{j_{1}}-\hat{\sigma}^{3}
_{i_{1}}\big)\,.
\end{eqnarray}
When applying this  five-fold sequence of transformations to states
in $\mathbb{V}_{DFS_{2_{i}}}\otimes\mathbb{V}_{DFS_{2_{j}}}$, the
product $\hat{\sigma}^{3}_{i_{2}}\otimes\hat{\sigma}^{3}_{i_{1}}$ on
the right-hand side can be ignored, since it is proportional to the
logical identity operator, and introduces thus nothing but a global
phase factor. This leaves us with
$\frac{1}{2}\hat{\sigma}^{3}_{i_{2}}\otimes\hat{\sigma}^{3}_{j_{1}}$,
which is equivalent to
$-\frac{1}{2}\hat{\sigma}^{3}_{L_{i}}\otimes\hat{\sigma}^{3}_{L_{j}}\equiv
-\frac{1}{2}\hat{H}_{L_{i}L_{j}}$  (with the non-symmetric choice
$\varepsilon_{i}=0$ and $\varepsilon_{j}=1$ in Eq.
(\ref{logPaulimat})).
\par  Also here only interactions between two physical qubits at a time are required, but the technique has the drawbacks that it requires more pulses and can be used only when pairs $i$ and $j$ experience the same phase fluctuations. Nevertheless, it constitutes an alternative to spin-dependent forces, specially when Ising-like interactions are not readily available, as it appears to be the case with clock states connected via dipole Raman transitions.
\section{Implementation on trapped-ions}
\label{SecIII}
\label{implementation}
\par We consider next $N$ pairs of ions confined in a linear Paul trap, or in an arrangement of multi-connected
linear Paul traps, where  individual laser addressing is available.
The collective vibrational mode along the axial direction $z$, of
frequency $\nu$, might be the center-of-mass or stretch mode. The $i$-th logical qubit is
encoded into a pair $i$ of neighboring ions $i_{1}$ and $i_{2}$. We
assume each ion $i_{n}$ ($n=1$ or $2$) to have a mass $M$ and an
equilibrium position $z_{0_{i_{n}}}$. The ions may either possess
three energy levels in a $\Lambda$ configuration: two long-lived
ground-state levels, and an excited electronic state; or two energy
levels, one of which is a metastable state, and the other the ground
state. In both cases, we label the physical qubit states as
$\ket{\uparrow_{i_{n}}}\equiv\ket{0_{i_{n}}}$ and
$\ket{\downarrow_{i_{n}}}\equiv\ket{1_{i_{n}}}$, and their internal
transition frequency $\omega_{0}$. For  three-level ions the
physical qubit states are encoded in the two long-lived ground-state
levels, $\omega_{0}$ is typically in the microwave region, and the
qubit states are typically connected by a dipole Raman
transition through the excited electronic state, driven by two laser
beams $A$ and $B$, of frequencies $\omega_A$ and $\omega_B$ and wave
vectors along the $z$ direction $k_{A_{z}}$ and $k_{B_{z}}$.
For two-level ions,  in turn, the metastable state
encodes $\ket{\uparrow_{i_{n}}}\equiv\ket{0_{i_{n}}}$, the ground
state $\ket{\downarrow_{i_{n}}}\equiv\ket{1_{i_{n}}}$, and they are
connected by a weak quadrupole optical transition directly driven by
a single laser $L$, of  frequency $\omega_L$ and wave vector along
the $z$ direction $k_{L_{z}}$. 
\subsection{Single-logical-qubit gates: $\hat{\sigma}^{3}_{L}$}
\label{3} We show first how to implement Hamiltonian
(\ref{Hamlogqubit}) for the case $B^{0}_{i}=B^{1}_{i}=B^{2}_{i}=0$.
In this case it suffices to induce an AC Stark shift on only one of
the members of the pair, for example ion $i_{n}$, which can be done
by the application of off-resonant fields $\delta$-detuned from the
carrier transition. The interaction Hamiltonian in the interaction
picture with respect to the unperturbed Hamiltonian without the
laser field, and in the rotating wave approximation (RWA), with the
condition $\omega_{0}\gg \nu \gg \delta$, then reads:
$\hat{H}_{i_{n}}=\hbar\Omega_{i_{n}}\hat{\sigma}^{+}_{i_{n}}e^{i[\delta
t+\varphi_{i_{n}}]}+h.c.$ Here  $\Omega_{i_{n}}$ is the effective
Rabi frequency coupling $\ket{\uparrow_{i_{n}}}$ with
$\ket{\downarrow_{i_{n}}}$ and $\varphi_{i_{n}}$ is the spin phase, the field's
effective optical phase at position $z_{0_{i_{n}}}$.
\par From now on we will always work in the dispersive regime $|\Omega_{i_{n}}|\ll\delta$, in which perturbative
calculations with $\frac{\Omega_{i_{n}}}{\delta}$ as a perturbation parameter are valid. In fact, a time-dependent
second-order
perturbative calculation, yields an effective time-independent Hamiltonian given by:
\begin{eqnarray}
\hat{H}_{i_{n}}=\hbar\frac{|\Omega_{i_{n}}|^{2}}{\delta}\hat{\sigma}^{3}_{i_{n}}\,
\label{sygma3}
\end{eqnarray}
Since, according to Eq.~(\ref{logPaulimat}), $\hat{\sigma}^{3}_{L_{i}}$ coincides with $\hat{\sigma}^{3}_{i_{1}}$ for
 the non-symmetric choice $\varepsilon_i=1$ and with $-\hat{\sigma}^{3}_{i_{2}}$ for $\varepsilon_i=0$, it is
\begin{equation}
\hat{H}_{i_{n}}=B_i^3\hat{\sigma}^3_{L_{i}}\,,
\label{implementlogsigma3}
\end{equation}
with $B^3_i\equiv\pm\hbar\frac{|\Omega_{i_{n}}|^{2}}{\delta}$, the ``$+$" (``$-$") sign corresponding to $n=1$ ($n=2$); implementing thus the desired logical Hamiltonian.
\par It is important to notice that in the above derivation, as well as in the rest of the paper, the resolved-sideband limit, $|\Omega_{i_{n}}|\ll\nu$, is assumed. In this regime, by tuning the laser frequency, it is always possible to select the stationary terms of the Hamiltonian and to neglect ---in the RWA--- all other terms rotating at the different vibrational modes' frequencies. This was exploited here to neglect terms involving any vibrational mode frequency by setting $\omega_L$  (or $\omega_A-\omega_B$) close to $\omega_0$, and is exploited in the next subsections to select the desired vibrational mode by setting it close to resonance with a sideband transition to such mode.
\subsection{Single-logical-qubit gates: $\hat{\sigma}^{\phi}_{L}$}
\label{phi}
\par We now concentrate on the implementation of Hamiltonian $\hat{H}_{i_{1}i_{2}}=C_{i}\hat{\sigma}^{\phi_{i}}_{L_{i}}$,
where $C_{i}$ is a constant and  $\hat{\sigma}^{\phi_{i}}_{L_{i}}$,
defined as
$\hat{\sigma}^{\phi_{i}}_{L_{i}}\equiv\cos(\phi_{i})\hat{\sigma}^{1}_{L_{i}}+
\sin(\phi_{i})\hat{\sigma}^{2}_{L_{i}}\equiv
e^{-i\phi_{i}}\hat{\sigma}^{+}_{L_{i}}+e^{i\phi_{i}}\hat{\sigma}^{-}_{L_{i}}$,
is the operator contained in the equatorial plane of the logical
Bloch sphere with azimuth angle $\phi_{i}$. This is equivalent to
Hamiltonian (\ref{Hamlogqubit}) with $B^{0}_{i}=B^{3}_{i}=0$,
$B^{1}_{i}\equiv C_{i}\cos(\phi_{i})$ and $B^{2}_{i}\equiv
C_{i}\sin(\phi_{i})$. For any fixed value of $\phi_{i}$ the ability
to implement such Hamiltonian, together with
Hamiltonian~(\ref{implementlogsigma3}), suffices to generate any
SU(2) operation on the $i$-th logical qubit.
\par In this case it is possible to use the SM-gate \cite{Sorensen-Molmer,Lee}, driven
by one field detuned by $\delta$ from the red sideband, plus another
one detuned by $-\delta$ from the blue one. Here we show  nonetheless
 that only one
of these fields suffices as long as one remains in
$\mathbb{V}_{DFS_{2_{i}}}\otimes\mathbb{V}_{DFS_{2_{j}}}$. We
extend the ideas of Ref.~\cite{schmidt-kaler} and consider a laser field
irradiating simultaneously both ions of the $i$-th pair. When the
laser frequency or laser frequency difference is close to resonance with a
sideband transition, a coupling between the internal qubit
states and the relevant vibrational mode is possible. We choose the
first red sideband transition for definiteness, but the blue one
would work just as well. That is, we set
$\omega_A-\omega_B=\omega_{0}-\nu-\delta$ and $\Delta k_{z}\equiv
k_{B_{z}}-k_{A_{z}}\neq 0$  (non-copropagating beams is a further requirement for Raman couplings), or $\omega_L=\omega_{0}-\nu-\delta$. All
other vibrational modes can be neglected under the RWA because we
are in the resolved-sideband limit and they give no stationary
contribution. The Lamb-Dicke parameter is defined as
$\eta_{\nu}\equiv\Delta
k_{z}\frac{z_{\nu}}{\sqrt{2N}}\equiv\frac{1}{\sqrt{2N}}\Delta
k_{z}\sqrt{\frac{\hbar}{M\nu}}$, or $\eta_{\nu}\equiv k_{L_{z}}
\frac{z_{\nu}}{\sqrt{2N}}\equiv\frac{1}{\sqrt{2N}}
k_{L_{z}}\sqrt{\frac{\hbar}{M\nu}}$, where $z_{\nu}$ is the
root-mean-square width of the motional ground-state wave packet. We
assume next that the system is in the Lamb-Dicke limit (LDL)
$\eta_{\nu}^{2}(n_{\nu}+1/2)\ll 1$, with $n_{\nu}$ the mean phonon
population, meaning that the wave packet is very localized as
compared to the fields' wavelengths $2\pi \Delta k_{z}^{-1}$ or
$2\pi k_{L_{z}}^{-1}$.  In this case the interaction Hamiltonian in
the RWA  is given by
$\hat{H}_{i_{1}i_{2}}=\hbar[\Omega_{i_{1}}\hat{\sigma}^{+}_{i_{1}}(e^{i\nu
t}+i\eta_{\nu}\hat{a}_{\nu})e^{i(\delta
t+\varphi_{i_{1}})}+\Omega_{i_{2}}\hat{\sigma}^{+}_{i_{2}}(e^{i\nu
t}-i\eta_{\nu}\hat{a}_{\nu})e^{i(\delta t+\varphi_{i_{2}})}+ {\rm h.
c.}]$, where $\hat{a}_{\nu}$ is the annihilation operator of one
phonon. Notice that here, in spite of being in the resolved sideband
limit, we have not neglected the fast oscillating term proportional
to $e^{i\nu t}$, since for very low values of $\eta_{\nu}$ the
contribution of the latter might be comparable to that of the
stationary term proportional to $\eta_{\nu}$.
\par Taking both Rabi frequencies equal,
$\Omega_{i_{1}}=\Omega_{i_{2}}\equiv\Omega_{i}$, yields the time-independent effective Hamiltonian:
\begin{widetext}
\begin{eqnarray}
  \hat{H}_{i_{1}i_{2}}=\hbar\frac{|\Omega_{i}|^{2}}{\nu+\delta}(\hat{\sigma}^{z}_{i_{1}}+\hat{\sigma}^{z}_{i_{2}})+
  \hbar\frac{|\Omega_{i}\eta_{\nu}|^{2}}{\delta}\Big[\hat{I}+(\hat{\sigma}^{z}_{i_{1}}+\hat{\sigma
}^{z}_{i_{2}})(\hat{n}_{\nu}+1/2)
  -(\hat{\sigma}^{+}_{i_{1}}\otimes\hat{\sigma}^{-}_{i_{2}}e^{i(\varphi_{i_{1}}-\varphi_{i_{2}})}+ {\rm h. c.})\Big],
\label{Heff}
\end{eqnarray}
with $\hat{n}_{\nu}\equiv\hat{a}^{\dagger}_{\nu}\hat{a}_{\nu}$.
The identity operator $\hat{I}$ can be omitted as it only generates an irrelevant global phase factor; and so can the terms proportional $\hat{\sigma}^{z}_{i_{1}}+\hat{\sigma}^{z}_{i_{2}}$, for they are equivalent to $\hat{0}_{L_{i}}$ (taking $\rho_{i}=\zeta_{i}=\vartheta_{i}=\theta_{i}=\kappa_{i}=\lambda_{i}=\varsigma_{i}=0$ in Eq. (\ref{lognull})). We thus see that in $\mathbb{V}_{DFS_{2_{i}}}$ Hamiltonian (\ref{Heff}) is equivalent to
\begin{eqnarray}
\label{Heff2}
\nonumber
&\hat{H}_{i_{1}i_{2}}=\\
-&\hbar\frac{|\Omega_{i}\eta_{\nu}|^{2}}{\delta}\big[\hat{\sigma}^{+}_{i_{1}}\otimes\hat{\sigma}^{-}_{i_{2}}e^{i\phi_{i}}+ {\rm h. c.}\big]=
-\hbar\frac{|\Omega_{i}\eta_{\nu}|^{2}}{2\delta}\big\{\cos(\phi_{i})\big[\hat{\sigma}^{1}_{i_{1}}\otimes\hat{\sigma}^{1}_{i_{2}}+\hat{\sigma}^{2}_{i_{1}}\otimes\hat{\sigma}^{2}_{i_{2}}\big]
+\sin(\phi_{i})\big[\hat{\sigma}^{2}_{i_{1}}\otimes\hat{\sigma}^{1}_{i_{2}}-\hat{\sigma}^{1}_{i_{1}}\otimes\hat{\sigma}^{2}_{i_{2}}\big]\big\}\,,
\end{eqnarray}
\end{widetext}
with $\phi_{i}\equiv\varphi_{i_{1}}-\varphi_{i_{2}}$. A direct exchange of quanta between both ions through a virtual excitation of the vibrational mode. It is in turn immediate to express (\ref{Heff2}) as the desired Hamiltonian
\begin{equation}
\hat{H}_{i_{1}i_{2}}=C_{i}\hat{\sigma}^{\phi_{i}}_{L_{i}}\, ,
\label{desired}
\end{equation}
with $C_{i}\equiv-\hbar\frac{|\Omega_{i}\eta_{\nu}|^{2}}{\delta}$, and where $\beta_i=\gamma_i=1/2$ have been taken in  Eq.~(\ref{logPaulimat}).
\subsection{Two logical-qubit gates}
\label{Universal}
\par  A $\hat{\sigma}^{3}\otimes\hat{\sigma}^{3}$-type interaction between physical qubits from different pairs is
required to realize Hamiltonian (\ref{UniqueHamiltonian}). However,
interaction schemes such as the one described in the previous
subsection that use the vibrational mode as a virtual mediator
always involve products as
$\hat{\sigma}^{\pm}\otimes\hat{\sigma}^{\pm}$. So a
$\hat{\sigma}^{3}\otimes\hat{\sigma}^{3}$ effective interaction,
with no explicit dependence on $\hat{a}_{\nu}$ or
$\hat{a}_{\nu}^{\dagger}$, appears only as a fourth-order
contribution, negligible as compared to the contributions from
previous orders. Therefore, it is very ineffecient to
realize a non-local gate between two logical qubits using only
two-body interactions, with no explicit dependence on the
vibrational operators, under the requirement that the states
involved in the operation stay in the encoded subspace
$\mathbb{V}_{DFS_{2_{i}}}\otimes\mathbb{V}_{DFS_{2_{j}}}$. If, on
the other hand, the vibrational mode is allowed to be actually
populated, instead of just being used as a virtual mediator, optical
forces that exert a state-dependent force onto the ions can be used
to generate effectively such an interaction \cite{Leibfried}.
\par   As the implementation of these optical-forces is described elsewhere  \cite{Leibfried,Lee,Blinov,Kihwan, Leandro}, we show here how to implement
the alternative encoded re-coupling scheme. This requires $\hat{\sigma}^{+}\otimes\hat{\sigma}^{-}$-type interactions between different physical qubits
from both pairs to realize the sequence of transformations in (\ref{re-coupling}). Such sequence of pulses momentarily takes the states out of 
$\mathbb{V}_{DFS_{2_{i}}}\otimes\mathbb{V}_{DFS_{2_{j}}}$ but never out of $\mathbb{V}_{DFS_{4_{ij}}}$. This implies that the bichromatic gate described in the previous subsection cannot be used here, since the two terms proportional to $\hat{\sigma}^{z}_{i_{1}}+\hat{\sigma}^{z}_{i_{2}}$ eliminated from Hamiltonian~(\ref{Heff}) because of being proportional to $\hat{0}_{L_{i}}$ do have a finite support on $\mathbb{V}_{DFS_{4_{ij}}}$. Hamiltonian (\ref{Heff}) is not equivalent to (\ref{Heff2}) outside $\mathbb{V}_{DFS_{2_{i}}}\otimes\mathbb{V}_{DFS_{2_{j}}}$.
\par  The first term in (\ref{Heff}) is only a Stark shift that contains no interaction between both physical qubits. From the formal point of view
one could simply leave it in the evolution and then undo its action at the end by just applying local pulses. An experimentally-accessible approach 
to compensate for it is, either to perform an effective qubit frequency renormalization \cite{Lee}, or to use compensation-laser techniques
\cite{Haeffner03}. We therefore disregard it, which leaves us with
\begin{eqnarray}
\label{Heff3}
\hat{H}_{i_{1}i_{2}}=\hbar\frac{|\Omega_{i}\eta_{\nu}|^{2}}{\delta}\big[\hat{I}+(\hat{\sigma}^{z}_{i_{1}}+\hat{\sigma }^{z}_{i_{2}})(\hat{n}_{\nu}+1/2)
\nonumber\\
-(\hat{\sigma}^{+}_{i_{1}}\otimes\hat{\sigma}^{-}_{i_{2}}e^{i\phi_{i}}+ {\rm h. c.})\big]\, ,
\end{eqnarray}
still containing the second term proportional to
$\hat{n}_{\nu}+1/2$. This term entangles the
internal and motional degrees of freedom, so that unless the system
is previously cooled to, and kept in, its motional ground state
$n_{\nu}=0$, it makes the action of the gate explicitly dependent on
the vibrational state. In order to circumvent this we notice that adding Hamiltonian (\ref{Heff3}) to
itself, with the replacements $\delta\leftrightarrow-\delta$ and
$\phi_{i}\leftrightarrow\phi_{i}+\pi$, yields exactly twice Hamiltonian (\ref{Heff2}) without an
explicit dependence on the vibrational operators, even when applied
to any two ions from different pairs and for states outside
$\mathbb{V}_{DFS_{2_{i}}}\otimes\mathbb{V}_{DFS_{2_{j}}}$.
\par Now, the latter is exactly the effective Hamiltonian of the system when, simultaneously with the field so far considered, a second field is applied on both ions. Since the two ions can now be any ions from any pair we drop the subindex $i$. The second field (herein labeled with a ``$\sim$") must have the same base Rabi frequency $\tilde{\Omega}=\Omega$ and Lamb-Dicke parameter $\tilde{\eta}_{\nu}=\eta_{\nu}$ as the first one;  must be exactly $\pi$ radians out of phase with it, $\tilde{\phi}=\phi+\pi$; and must be $-\delta$-detuned from the same sideband transition: $\omega_{\tilde{L}}=\omega_{0}-\nu+\delta$.
\par This bichromatic scheme differs from the SM-gate in an important way: the latter is based on Raman beams detuned from opposite sidebands rather
than the same sideband, as proposed here. The SM-gate can operate in both the dispersive regime $|\Omega_{k_{n}}|\ll\delta$, in which the vibrational
degree of freedom is also only used as a virtual mediator, and the ``fast'' regime of small $\delta$ --more naturally described as a
$\hat{\sigma}^{\phi}$-dependent force-- in which the motional degree of freedom is actually populated during the gate evolution. Nevertheless,
in both regimes the SM-gate Hamiltonian includes terms as $\hat{\sigma}^{+}\otimes\hat{\sigma}^{+}$ and $\hat{\sigma}^{-}\otimes\hat{\sigma}^{-}$,
which are undesired in this context for, even though they do not have support on $\mathbb{V}_{DFS_{2_{i}}}\otimes\mathbb{V}_{DFS_{2_{j}}}$, they take some states out of $\mathbb{V}_{DFS_{4_{ij}}}$. The SM-gate is therefore not useful for the implementation of the encoded re-coupling scheme.
\par Since the encoded re-coupling scheme involves several pulses, the duration of the procedure must be compared to realistic entanglement lifetimes. For instance, taking the experimental values at the Innsbruck experiment \cite{Haffner2}: $\Omega=2\pi\times 100$ kHz, $\eta_{\nu}=0.0165$ and $\delta=2\pi\times 16.5$ kHz (10 $\times\eta_{\nu} \Omega$ ), the time required to realize, for instance, the pulse $e^{-i[\hat{\sigma}^{+}\otimes\hat{\sigma}^{-}+ {\rm h.c.}]\pi/2}$ is $\tau\equiv\pi\delta/2(\Omega\eta_{\nu})^2=3$~ms, which is four orders of magnitude smaller than the $20$ seconds robust entanglement reported there. We also note that in the case of Raman transitions the effective Lamb-Dicke parameter $\eta_{\nu}$ is typically larger, yielding a considerable speed-up.
\begin{figure}[t]
\begin{center}
\includegraphics[width=0.95\linewidth]{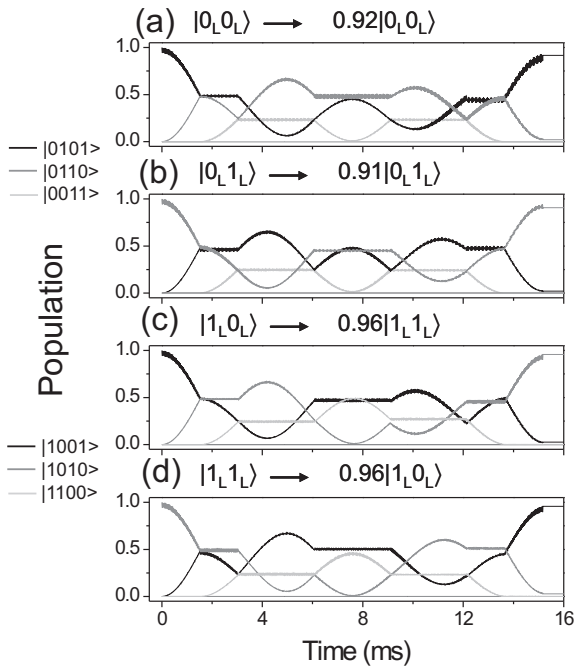}
 \caption{
 \label{CNOT}
State evolution of two logical qubits encoded in 4 physical qubits under the
action of a CNOT gate implemented with the encoded re-coupling scheme for four different input states. While in (a) and (b) mainly the physical states
$|$0101$\rangle$, $|$0110$\rangle$, $|$0011$\rangle$ are populated, in (c) and (d) mostly $|$1001$\rangle$, $|$1010$\rangle$,
 $|$1100$\rangle$ are involved. The presence of off-resonant excitations leads to very rapid micro-oscillations in the populations that, in
this resolution, simply appear as thicker lines. Pulse shaping can be used to suppress the imperfections induced by those off-resonant
excitations.}
\end{center}
\end{figure}
\par In addition, we have numerically simulated the pulse sequence~(\ref{re-coupling})
to generate a logical $\pi$-phase gate. The model used for the
simulation is that of the  usual Jaynes-Cummings  Hamiltonian only
under the optical RWA, and where the terms in its Taylor expansion
of order higher than 2 in the Lamb-Dicke parameter where neglected.
Two $\pi/2$-pulses on the logical qubits were inserted just before
and after the phase gate to turn it into a logical CNOT gate that,
written in terms of the physical-qubit states, has the following
truth table:
\begin{eqnarray}
\label{CNOTgate}
|0_{i_{1}}1_{i_{2}},0_{j_{1}}1_{j_{2}}\rangle &\longrightarrow&
|0_{i_{1}}1_{i_{2}},0_{j_{1}}1_{j_{2}}\rangle ,\nonumber \\
|0_{i_{1}}1_{i_{2}},1_{j_{1}}0_{j_{2}}\rangle &\longrightarrow&
|0_{i_{1}}1_{i_{2}},1_{j_{1}}0_{j_{2}}\rangle ,\nonumber \\
|1_{i_{1}}0_{i_{2}},0_{j_{1}}1_{j_{2}}\rangle &\longrightarrow&
|1_{i_{1}}0_{i_{2}},1_{j_{1}}0_{j_{2}}\rangle ,\nonumber\\
|1_{i_{1}}0_{i_{2}},1_{j_{1}}0_{j_{2}}\rangle &\longrightarrow&
|1_{i_{1}}0_{i_{2}},0_{j_{1}}1_{j_{2}}\rangle .
\end{eqnarray}
Fig.~\ref{CNOT} shows the numerically calculated evolution of
selected populations during the CNOT operation. The total required
pulse area is 5$\pi$ and thus the total required time for a CNOT is
approximately 15~ms. For all four test cases fidelities exceeding
90\% are calculated. We used the values from above for the laser
settings. The motional mode frequency was chosen to be 2$\pi \times$
1.2 MHz, a typical value in the Innsbruck experiments. Decoherence
effects such as magnetic field fluctuations and laser frequency
fluctuations are not considered because the evolution takes place
predominantly in the DFS. The assumed addressing error of 5\% on
adjacent ions  reduces the fidelities by about 3\% and off-resonant
excitations produce a 3\% error. Other decoherence sources, like
intensity fluctuations, motional heating, etc., are expected to
contribute not significantly. The errors due to off-resonant
excitations can be greatly reduced by pulse shaping, i.e. switching
laser pulses adiabatically as compared to the Rabi-frequencies.
Addressing errors can be reduced considerably with composite pulse
sequences such that they appear only in second order. Thus we
estimate that the total infidelities of the proposed scheme is
potentially well below 1$\%$ even with present technology, so that
the gate fulfills  the requirements set  in \cite{Knill} for
fault-tolerant quantum computation. We note, however, that for
useful quantum computation, higher gate fidelities than estimated
here reduce the overhead dramatically.
\subsection{Phase sensitivity}
\label{Phase}
\par Let us briefly discuss on the sensitivity of the protocol to fluctuations of the optical phase of the driving fields
due to relative path instabilities, which can be a serious limiting
factor for the fidelity of the gates \cite{Lee, Sackett}. For the
implementation of single qubit operations in the DFS (Hamiltonians~(\ref{implementlogsigma3}) and
(\ref{desired})) on qubits using optical transitions this will not represent a major problem, since
co-propagating laser beams can be used and thus relative phase fluctuations can be made quite small. A very similar situation arises for Raman-driven
qubits as each pair of non-copropagating laser beams acts simultaneously on neighboring ions and can be viewed as one effective field,
with phase fluctuations between both ions small as in the above case. 
We thus conclude that path length differences can be efficiently controlled in the single logical qubit case.
\par We now turn to the two-logical-qubit gates. In contrast to single qubit operations, here each bichromatic
beam acts simultaneously on two ions that are not necessarily neighbors. Nevertheless, even for ion-spacings of up to 1mm the beams take essentially
the same path and thus e.g. relative fluctuations of the air's refraction index are not significant. Furthermore, since the five-fold pulse
sequence~(\ref{re-coupling}) yields a $\pi$-phase gate, which does not
depend on the spin phase, interferometric stability is
required only throughout the pulse sequence. Therefore long-term
interferometric stability is not necessary.
\section{Conclusion}
\label{conclusion}
We considered the different interactions involving only two physical qubits at a time that generate
universal quantum gates on collective-dephasing-free-encoded qubits, and described feasible experimental demonstrations of each of
them with trapped-ions using existing technology.
A general formal classification of all two-body dynamics able to generate such gates without the system's
state ever leaving the encoded subspace was provided in terms of the Pauli operators associated to each physical
qubit, together with the explicit presentation of the allowed Hamiltonians. The implementation of these Hamiltonians
operates in the individual laser addressing regime and requires no ion-shuttling, so that it complements the collective-ion-addressing
based proposals. Also, no ground-state cooling is needed, provided that the ions always remain in the Lamb-Dicke regime. In addition, it makes use of a novel two-ion gate based on bichromatic
Raman fields that can be applied to clock states connected via dipole Raman transitions. Finally, even though this gate is particularly
well-suited for implementing universal quantum computing in DFS's, it constitutes by itself a potentially useful alternative to other entangling gates
outside the context of DFS's.
\begin{acknowledgments}
The authors thank Nicim Zagury for very fruitful conversations and
Andreas Buchleittner for the hospitality  in Dresden, where the
first discussions that led to this work took place. We gratefully
acknowledge support by CNPq, FAPERJ, CAPES, the brazilian Millennium
Institute for Quantum Information, the Austrian Science Fund (FWF),
the European Commission (CONQUEST, SCALA networks) and the
Institut f\"ur Quanteninformation GmbH. K. Kim acknowledges
funding by the Lise-Meitner program of the FWF.
\end{acknowledgments}


\begin{thebibliography}{99}
\bibitem{Shor-Machiavello-Steane-Gottesman-Calderbank-Miquel}
P. W. Shor, Phys. Rev. A {\bf 52}, R2493 (1995); A. Ekert and C.
Machiavello, Phys. Rev. Lett {\bf 77}, 2585 (1996); A. Steane, Phys.
Rev. Lett. {\bf 77}, 793-797 (1996); D. Gottesman, Phys. Rev. A {\bf 54}, 1862 (1996); A. R. Calderbank,  E. M. Rains, P.
M. Shor, and N. J. Sloane,
Phys. Rev. Lett {\bf 78}, 405 (1997); R.
Laflamme, C. Miquel, J. P. Paz and W. H. Zurek
Phys. Rev. Lett {\bf 77}, 198 (1996).
\bibitem{Palma-Barenco-Zanardi}
G. M. Palma, K. A. Suominen, and A. K. Ekert,
Proc. R. Soc. London Sect. A {\bf 452}, 567 (1996); A. Barenco,  A. Berthiaume, D. Deutsch, A. K. Ekert, R. Jozsa, and C. Machiavello,
{\bf SIAM J.} Comp. {\bf 26}, 1541 (1997);  P. Zanardi and
M. Rasetti, Phys. Rev. Lett. {\bf 79}, 3306 (1997).
\bibitem{Lidar-Bacon}
D. A. Lidar,  D. Bacon, K. B. Whaley, Phys. Rev. Lett. {\bf 82} ,
4556 (1999); D. Bacon, J. Kempe, D. A. Lidar, and K. B. Whaley,
Phys. Rev. Lett. {\bf 85}, 1758 (2000).
\bibitem{Bacon}
D. Bacon, quant-ph/0305025.
\bibitem{Duan-Zanardi-Lidar1}
L. M. Duan and G. C. Guo, Phys.
Rev. Lett {\bf 79}, 1953 (1997); P. Zanardi and M. Rasetti, {\it ibid. }  {\bf 79}, 3306 (1997); D. A. Lidar,  I. L. Chuang, and K. B. Whaley,
Phys. Rev. Lett. {\bf 81}, 2594 (1998).
\bibitem{Kielpinsky2}
D. Kielpinski, C. Monroe, and D. J. Wineland, Nature {\bf 417}, 709 (2002).
\bibitem{Kwiat}
P. G. Kwiat, A. J. Berglund, J. B. Altepeter, and A. G. White
Science {\bf 290}, 498 (2000).
\bibitem{Fortunato-Viola}
L. Viola, E. M. Fortunato, M. A. Pravia, E. Knill, R. Laflamme,
D. G. Cory, Science {\bf 293}, 2059 (2001); E. M. Fortunato, L. Viola, J. Hodges, G. Teklemariam, and D. G. Cory, New J. Phys.
{\bf 4}, 5 (2002).
\bibitem{Kielpinsky}
D. Kielpinski, V. Meyer, M. A. Rowe, C. A. Sackett, W. M. Itano, C. Monroe, and D.
J.
Wineland, Science {\bf 291}, 1013 (2001).
\bibitem{Roos}
C. F. Roos, G. P. T. Lancaster, M. Riebe, H. H\"{a}ffner, W. H\"{a}nsel, S. Gulde,
C.
Becher, J. Eschner, F. Schmidt-Kaler, and R. Blatt, Phys. Rev. Lett. {\bf 92},
220402
(2004).
\bibitem{Langer}
C. Langer, R. Ozeri, J. D. Jost, J. Chiaverini, B. DiMarco, A. Ben-Kish, R. B.
Blakestad, J. Britton, D. B. Hume, W. M. Itano, D. Liebfried, R. Riechle, T.
Rosenband,
T. Schaetz, P. O. Schmidt, and D. J. Wineland, Phys. Rev. Lett. {\bf 95}, 060502
(2005).
\bibitem{Haffner2}
H. H\"{a}ffner, F. Schmidt-Kaler, W. H\"{a}nsel, C. F. Roos, T.
K\"{o}rber, M.Chwalla, M. Riebe, J. Benhelm, U. D. Rapol, C. Becher,
and R. Blatt, Appl. Phys. B {\bf 81}, 151 (2005).
\bibitem{Wu-Zanardi-Lidar}
D. A. Lidar and L. A. Wu, Phys. Rev. A {\bf 67}, 032313 (2003); L. A. Wu, P. Zanardi and D. A. Lidar, Phys. Rev. Lett. {\bf 95}, 130501 (2005).
\bibitem{Sorensen-Molmer}
A. S\o rensen and K. M\o lmer, Phys. Rev. Lett. {\bf 82}, 1971
(1999);  A. S\o rensen and K. M\o lmer, Phys. Rev. A {\bf 62},
022311 (2000).
\bibitem{Sackett}
C. A. Sackett, D. Kielpinski, B. E. King, C.
Langer, V. Meyer, C. J. Myatt, M. Rowe, Q. A. Turchette, W. M.
Itano, E. J. Wineland, and C. Monroe, Nature \bf 404\rm, 256 (2000).
\bibitem{Duan}
L.-M. Duan, Phys. Rev. Lett. {\bf 93}, 100502 (2004).
\bibitem{Cen-Wang-Wang}
L.-X. Cen, Z. D. Wang, S. J. Wang, Phys. Rev. A {\bf 74}, 032321 (2006).
\bibitem{Leibfried}
D. Leibfried, B. DeMarco, V.
Meyer, D. Lucas, M. Barret, J. Britton, W. M. Itano, B
Jelenkovi\'{c}, C. Langer, T. Roseband, D. J. Wineland, Nature {\bf
422}, 412 (2003).
\bibitem{Lee}
P. J. Lee, K-A. Brickman, L. Deslauriers, P. C. Haljan, L.-M. Duan, C.
Monroe, J. Opt. B: Quantum. Semiclass. Opt. {\bf 7}, 371 (2005).
\bibitem{Blinov}
B. Blinov, D. Leibfried, C. Monroe, D. J. Wineland, Quantum Inf. Process. {\bf 3}, 1 (2004).
\bibitem{Haljan}
P. C. Haljan, K-A. Brickman, L. Deslauriers, P. J. Lee, C.
Monroe, Phys. Rev. Lett.  {\bf 94}, 153602 (2005).
\bibitem{Clock-states}
J.J. Bollinger, D.J. Heinzen, W.M. Itano, S.L. Gilbert, and D.J. Wineland, IEEE Trans. Instrum. Meas. {\bf 40}, 126 (1991);
P. T. H. Fisk, M. J. Sellars, M. A. Lawn, C. Coles, A. G. Mann, D. G. Blair, IEEE Trans. Instrum. Meas. {\bf 44}, 113 (1995).
\bibitem{Kihwan}
K. Kim, et al, in preparation (2007).
\bibitem{Leandro}
L. Aolita, K. Kim, J. Benhelm, C. F. Roos, H. H\"affner, in preparation (2007).
\bibitem{Lidar-Wu3}
D. A. Lidar and L. A. Wu, Phys. Rev. Lett. {\bf 88}, 017905 (2001).
\bibitem{Haeffner03}
H. H\"{a}ffner, S. Gulde, M. Riebe, G. Lancaster, C. Becher, J.
Eschner, F. Schmidt-Kaler, and R. Blatt, Phys. Rev. Lett. {\bf 90},
143602 (2003).
\bibitem{schmidt-kaler}
F. Schmidt-Kaler, H. H\"affner, S. Gulde, M. Riebe, G. Lancaster, J. Eschner, C.
Becher and R. Blatt, Europhys. Lett. {\bf 65}, 587 (2004).
\bibitem{Knill}
E. Knill, Nature (London) 434, 39 (2005).
\end{thebibliography}
\end{document}